\documentclass[aps,prl,twocolumn,preprintnumbers,amsmath,amssymb,superscriptaddress]{revtex4-2}
\usepackage{graphicx}
\usepackage[caption=false]{subfig}
\usepackage{epsfig}
\usepackage{dcolumn}
\usepackage{bm}
\usepackage{color}
\usepackage{float}
\usepackage[colorlinks]{hyperref}
\usepackage{verbatim}
\usepackage{algorithm}
\usepackage{algorithmic}
\usepackage{tikz}
\usetikzlibrary{quantikz}

\def\bea{\begin{eqnarray}}
\def\eea{\end{eqnarray}}
\def\nn{\nonumber}
\def\ba{\begin{array}}
\def\ea{\end{array}}
\def\nn{\nonumber}

\def\ii{\mathfrak{i}}

\hypersetup{citecolor = blue}

\def\be{\begin{equation}}       \def\ee{\end{equation}}
\def\bea{\begin{eqnarray}}      \def\eea{\end{eqnarray}}
\def\ba{\begin{array}}
\def\ea{\end{array}}
\def\bnum{\begin{enumerate} }
\def\enum{\end{enumerate}}

\def\nn{\nonumber}

\def\=>{\Rightarrow}
\def\>{\rightarrow}

\def\eye2{Fathbb{I}}

\renewcommand{\v}[1]{{\bf #1}}

\renewcommand{\>}{\rangle}

\usepackage{listings}
\usepackage{color}
\definecolor{lightgray}{gray}{1}

\begin{document}

\title{Quantum electrodynamics under a quench}

\author{Ming-Rui Li}
\affiliation{Institute for Advanced Study, Tsinghua University, Beijing 100084, China}
\author{Shao-Kai Jian}
\email{sjian@tulane.edu}
\affiliation{Department of Physics and Engineering Physics, Tulane University, New Orleans, Louisiana, 70118, USA}

\date{\today}% It is always \today, today,

\begin{abstract}

Quantum electrodynamics (QED) is a cornerstone of particle physics and also finds diverse applications in condensed matter systems. 
Despite its significance, the dynamics of quantum electrodynamics under a quantum quench remains inadequately explored. 
In this paper, we investigate the nonequilibrium regime of quantum electrodynamics following a global quantum quench. 
Specifically, a massive Dirac fermion is quenched to a gapless state with an interaction with gauge bosons. 
In stark contrast to equilibrium (3+1)-dimensional QED with gapless Dirac fermions, where the coupling is marginally irrelevant, we identify a nonequilibrium fixed point characterized by non-Fermi liquid behavior. 
Notably, the anomalous dimension at this fixed point varies with the initial quench parameter, suggesting an interesting quantum memory effect in a strongly interacting system. 
Additionally, we propose distinctive experimental signatures for nonequilibrium quantum electrodynamics. 
\end{abstract}

\maketitle

%\tableofcontents
{\it Introduction.---} 
In recent years, the progress of experiment platforms like the ion trap systems~\cite{RevModPhys.83.863,Blatt2012,PRXQuantum.2.020343} and cold atom systems~\cite{Bernien2017,Ebadi2021} has provided people with programmable and highly coherent many-body quantum systems with abundant exotic physics. 
These developments have also facilitated the exploration of nonequilibrium physics in correlated quantum systems~\cite{Noel2022,Chertkov2023,PhysRevLett.124.170602,doi:10.1126/science.abg2530,PhysRevLett.129.243202}. 
In addition, pump-probe spectroscopy of correlated materials~\cite{doi:10.1126/science.1217423,Mankowsky2014} provides methods to drive the quantum many-body system far away from equilibrium, such as performing a sudden quench to the Hamiltonian parameters~\cite{RevModPhys.83.863,Eisert2015}. 
The postquench quantum dynamics can exhibit fruitful non-trivial behaviors that are distinguished from the well-studied equilibrium systems. 
It is well known that an isolated quantum many-body system that satisfies the eigenstate thermalization hypothesis~\cite{PhysRevE.50.888,PhysRevA.43.2046} (ETH) is expected to thermalize, and lose the memory of its initial state~\cite{RevModPhys.83.863,Rigol2008}. 
However, additional universal initial information other than the conserved quantities can be preserved for a long time in some exotic prethermal states~\cite{PhysRevLett.93.142002,doi:10.1126/science.1224953,langen2016prethermalization,Langen2013,Eigen2018,doi:10.1146/annurev-conmatphys-031016-025451,mori2018thermalization,PhysRevLett.123.170606,PhysRevB.103.125116,PhysRevLett.128.020601,doi:10.1126/science.abg8102,PhysRevLett.130.130401,doi:10.1126/science.abb4928,shu2023equilibration}. 
One such exotic phenomenon is the quantum memory phenomena~\cite{calabrese2005ageing,tauber2014field} which is widespread in different areas of physics. 
To understand these non-thermal behaviors, various cases have been studied and analyzed, including the proximity to integratibility~\cite{PhysRevB.84.054304,van2013relaxation,PhysRevLett.111.197203,smith2013prethermalization,PhysRevLett.115.180601,PhysRevA.94.013601,PhysRevB.101.094308}, non-thermal fixed point~\cite{PhysRevLett.101.041603,PhysRevB.84.020506,PhysRevA.86.013624,PhysRevD.89.074011,PhysRevD.92.025041,erne2018universal,prufer2018observation,PhysRevLett.122.170404,PhysRevLett.122.173001,PhysRevLett.122.122301,PhysRevLett.125.230601,claassen2021flow,marino2022dynamical}, quenched induced dynamical phase transition~\cite{PhysRevLett.96.136801,calabrese2007quantum,PhysRevLett.103.056403,PhysRevLett.105.220401,PhysRevLett.105.076401,kitagawa2011dynamics,PhysRevB.88.165115,PhysRevLett.110.136404,PhysRevB.88.201110,PhysRevLett.110.135704,karrasch2013dynamical,PhysRevB.88.024306,PhysRevB.91.205136,PhysRevLett.124.043001,muniz2020exploring,xu2020probing}.

Gapless Dirac fermion is a basic description of the low-energy excitations in various equilibrium condensed matter systems including Dirac semimetals, graphene, and surface modes of the topological insulators~\cite{RevModPhys.82.3045, RevModPhys.90.015001}. 
Though exotic quantum phase transitions have been explored for the equilibrium Dirac fermions, the dynamical phase transition remains mysterious. 
A previous study has revealed the universal prethermal dynamics of a Dirac system coupled to a bosonic field by Yukawa coupling~\cite{PhysRevLett.123.170606}. 
%They found the prethermal dynamics are controlled by two fixed points depending on the quench parameter, i.e., a usual equilibrium chiral Ising fixed point and a dynamical chiral Ising fixed point. 
%The non-trivial initial slip exponent is also studied there. 
Their study identified the unique non-equilibrium behaviors of the Dirac gapless fermions. 
On the other hand, when coupled to gauge bosons, the Dirac fermion system realizes quantum electrodynamics (QED).
QED is one of the most successful theories in particle physics~\cite{peskin2018introduction}. 
Furthermore, it has found applications in condensed matter physics, including the study of high-temperature superconductors~\cite{PhysRevB.66.094504,PhysRevB.66.054535}, graphene~\cite{KATSNELSON20073,GIULIANI2012461,PhysRevD.94.114010,PhysRevX.8.031015,PhysRevD.98.065008}, Weyl semimetals~\cite{PhysRevD.86.045001,PhysRevB.92.125115,PhysRevB.107.035131}, and fractional quantum anomalous Hall system~\cite{PhysRevX.5.031027,song2023emergent}. 
However, the quenching dynamics of QED remains an open question.

In this letter, we investigate the prethermal and nonequilibrium behavior of QED. 
More explicitly, a Dirac fermion with $U(1)$ gauge bosons after a quench to a critical point, where the Dirac fermion becomes gapless, is studied. 
Using the Keldysh renormalization group~\cite{kamenev2023field} and calculating the perturbation up to the leading order, we find that, on a long timescale, the 2+1 and 3+1 dimensional QED exhibits fixed points distinguished from the equilibrium ones. 
In 3+1 dimensions, we identify a nontrivial fix point with non-Fermi liquid behaviors. 
At this nontrivial fixed point, the anomalous dimension of the Dirac fermion depends on the initial quench parameter, indicating a quantum memory effect.
This is in stark contrast to the equilibrium QED, in which the interaction is marginally irrelevant. 
We also propose the differential conductance as a unique experimental signature of such a non-Fermi liquid behavior. 
On the other hand, in the 2+1 dimensions, we find that the quenched QED eventually flows to a non-interacting theory, which is also different from the equilibrium QED$_3$. 
In the following, we focus on the investigation of 3+1 dimensional QED, and leave the discussion of 2+1 dimensional QED to the Appendix. 

%We further calculated the current response to external fields and showed that the memory effect can be observed by the current measurement.

\begin{figure}
	\includegraphics[width=8cm]{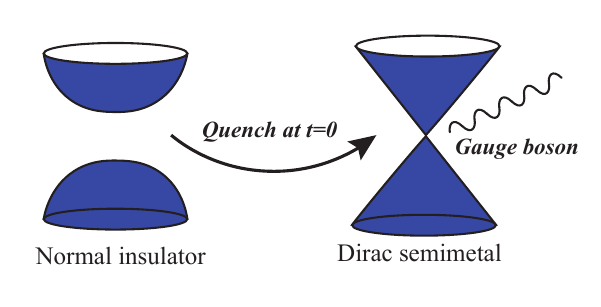}	\caption{A schematic plot of the quench from a normal insulator phase to a Dirac semimetal phase at time $t=0$. \label{fig:quench}}
\end{figure}

{\it Feynman propagators after a sudden quench.---} To study the prethermal behavior of quantum electrodynamics, we consider a quench of a free massive Dirac fermion with mass $m$ at temperature $\beta$ to a massless Dirac fermion at time zero that couples to a $U(1)$ gauge field as shown in the Fiq.~\ref{fig:quench}. 
The dynamics of free Dirac fermion in the imaginary-time evolution and real-time evolution are then described by the Hamiltonians
\bea \label{eq:dirac_free_hamiltonian}
	H_{E} =  \bm{k} \cdot \bm{\alpha} + m \gamma^0, \quad H_{R} =  \bm{k} \cdot \bm{ \alpha}, 
\eea
where the subscript $E$ and $R$ refer to the Euclidean evolution and real-time evolution, respectively. 
We will also call them prequench and postquench (free) Hamiltonian. 
${\bm k} \equiv (k_x, k_y, k_z)$ and $m$ are the momentum and mass of the Dirac fermion, respectively. Note that the postquench Dirac fermion is massless. 
$\gamma^\mu, \mu= 0,1,2,3$ denotes the Dirac matrix that satisfies $\{ \gamma^\mu, \gamma^\nu\} = 2  g^{\mu\nu}$. 
Here, we use the convention $g_{\mu\nu} = \text{diag}(1,-1,-1,-1)$. 
${\bm \alpha}$ is defined via ${\bm \alpha} \equiv (\gamma^0 \gamma^1, \gamma^0 \gamma^2, \gamma^0 \gamma^3)$. 
Such a quench protocol can be realized in numerous condensed matter systems.
For instance, it describes a sudden change of parameters that brings normal insulators to a Dirac semimetal phase, or describes a sudden quench to the normal insulator/topological insulator transition point.

\begin{figure}
	\includegraphics[width=6cm]{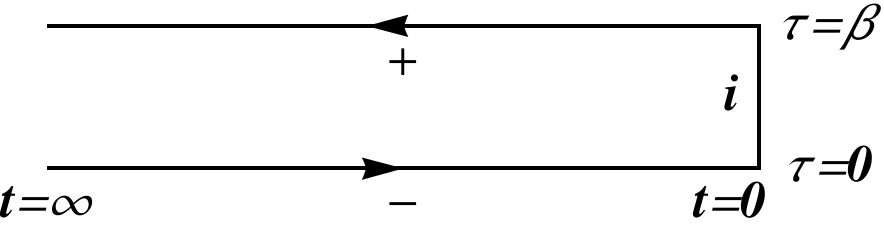} 
	\caption{A schematic contour of the quench dynamics. 
    The vertical (horizontal) line denotes the imaginary-time (real-time Keldysh) evolution.\label{fig:contour}}
\end{figure}

The Keldysh contour of the imaginary-time and the real-time evolution is illustrated in Fig.~\ref{fig:contour}. 
It is convenient to apply a conventional Keldysh rotation defined by
\begin{equation}
	\psi_{c/q} = \frac1{\sqrt2} (\psi_+ \pm \psi_-), \quad \psi_{c/q}^\dag = \frac1{\sqrt2} (\psi_+^\dag \mp \psi_-^\dag),
\end{equation}
and work in the Keldysh rotated field in the real-time evolution contour.
To obtain the Keldysh propagators, it is convenient to define the projector
\begin{equation}
\begin{aligned}
P_{\pm}(m)&= \frac12 \left( 1 \pm ( \bm{\hat k}_m \cdot \bm{\alpha} + \hat m \gamma^0) \right), \\
H_{E} P_{\pm}(m) &= \pm \xi P_{\pm}(m), \quad H_R  P_{\pm} = \pm |\bm{k}| P_{\pm},
\end{aligned}
\end{equation}
where $\bm{\hat k}_m = \xi^{-1} (k_x, k_y, k_z), \quad \hat m = \xi^{-1} m, \quad \xi \equiv \sqrt{\bm{k}^2 + m^2}$ and $P_\pm \equiv P_\pm(0)$. 
To proceed, we solve the equation of motion with the free Hamiltonian Eq.~\ref{eq:dirac_free_hamiltonian} in the Keldysh contour. Note that it can be done by solving the field variable and then matching the boundary conditions at $\tau = 0$ and $\tau = \beta $. For details, one can refer to the Appendix.
% We assume the value of $\psi$ at time $\tau =0$ to be $\psi_0$. 
% Then the solution in the imaginary-time contour is
% \bea
% 	\psi_i(\tau) = \left(e^{- \xi \tau} P_+(m) + e^{\xi \tau} P_-(m) \right) \psi_0.
% \eea

% On the other hand, the solution for real-time evolution is
% \begin{equation}
% \begin{aligned}
% 	\psi_c(t) &= \left( e^{-\ii |\bm{k}| t} P_+  + e^{\ii |\bm{k}| t} P_- \right) \psi_{c0},\\
%     \psi_q^\dag(t) &= \psi_{q0}^\dag \left(  e^{\ii |\bm{k}| t} P_+  + e^{-\ii |\bm{k}| t} P_-  \right),
% \end{aligned}
% \end{equation}
% with the boundary condition at time zero
% \bea
% 	\psi_{c0} = \frac1{\sqrt2 } (\psi(\beta) - \psi(0)), \quad \psi_{q0} = \frac1{\sqrt2 } (\psi(\beta) + \psi(0)).
% \eea 

With the knowledge of the solution, one can obtain the Keldysh propagator at the real-time contour that characterizes the quench protocol 
\begin{equation}
\begin{aligned}
	\langle \Psi(t_1) \bar\Psi(t_2)\rangle &= \ii \left( \ba{cccc} G_R & G_K \\ 0 &  G_A \ea  \right)(t_1, t_2), \\
	\ii G_K(t_1, t_2; \bm{k}) &= \tanh \frac{\beta \xi}2(e^{-\ii |\bm{k}|t_1} P_+ + e^{\ii |\bm{k}|t_1} P_-)\times\\
 &(P_+(m) - P_-(m)) (e^{\ii |\bm{k}|t_2} P_+ + e^{-\ii |\bm{k}|t_2} P_-) \gamma^0, \label{equ:keldyshpropagator} %\\
%	= \tanh \frac{\beta \xi}2 \left(\cos \chi (e^{-\ii |\bm{k}|(t_1-t_2)} P_+ - e^{\ii |\bm{k}|(t_1-t_2)} P_-)\gamma^0 + \sin \chi (e^{\ii |\bm{k}|(t_1+t_2)} P_+ + e^{-\ii |\bm{k}|(t_1+t_2)} P_-) \right)
\end{aligned}
\end{equation}
where $\Psi \equiv (\psi_c, \psi_q)^T$, $\bar\psi = \psi^\dag \gamma^0$. % and $\tan \chi = \frac{m}{|\bm k|}$. 
The quench parameter $m$ breaks the time translation symmetry of $G_K$, while other propagators are conventional 
\bea
	\ii G_{R/A}(t;\bm{k}) = \pm \theta(\pm t) \left( e^{-\ii |\bm{k}|t} P_+ + e^{\ii |\bm{k}|t} P_- \right) \gamma^0, %\\
%	\ii G_A(t;\bm{k}) = -\theta(-t) \left( e^{-\ii |\bm{k}|t} P_+ + e^{\ii |\bm{k}|t} P_- \right) \gamma^0.
\eea
where $\theta$ denotes the step function.

The $U(1)$ gauge fields coupled to the fermion can be described by the action:
\begin{equation}
\begin{aligned}
S&=\frac{1}{4} \int F_{\mu \nu} F^{\mu \nu}, %=-\frac{1}{2} \int A_\mu\left(\partial^2 g^{\mu \nu}-\partial^\mu \partial^v\right) A_v\\
%&=\frac{1}{2} \int A_\mu k^2 g^{\mu v} A_v.
\end{aligned}
\end{equation}
where $F_{\mu\nu} = \partial_\mu A_\nu - \partial_\nu A_\mu$ is the field strength tensor.
%The gauge coupling will be detailed in the next section. 
For later convenience, we can also apply the Keidysh rotation for gauge bosons:
\begin{equation}
    A_{c/q} = \frac1{\sqrt2} (A_+ \pm A_-).
\end{equation}
Then, after gauge fixing, we obtain the gauge boson propagator in the classical/quantum fields basis,
\begin{equation}
	\hat{D}_{\mu\nu}(t, \bm k) = \ii \left( \ba{cccc} D^K_{\mu\nu} & D^R_{\mu\nu} \\ D^A_{\mu\nu} &  0 \ea  \right)(t, \bm k),    
\end{equation}
with
\begin{equation}
\begin{aligned}
	 D_{\mu\nu}^R(t, \bm k) &= -\ii \langle A_{c,\mu}(t) A_{q,\nu}(0) \rangle =  \theta(t) g_{\mu\nu} \frac{\sin(|\bm{k}| t)}{ |\bm{k} | }, \\
	 D_{\mu\nu}^A(t, \bm k) &= -\ii \langle A_{q,\mu}(t) A_{c,\nu}(0) \rangle = -\theta(-t) g_{\mu\nu} \frac{\sin(|\bm{k}| t)}{ |\bm{k} | }, \\
	 D_{\mu\nu}^K(t, \bm k) &= -\ii \langle A_{c,\mu}(t) A_{c,\nu}(0) \rangle = \ii  g_{\mu\nu} \coth \frac{\beta |\bm{k}| }2 \frac{\cos(|\bm{k}| t) }{|\bm{k} |}.
\end{aligned}
\end{equation}

One can see that, since the information of the quench protocol is encoded in the fermion Keldysh propagator, only the Keldysh propagator~\ref{equ:keldyshpropagator} breaks time translation symmetry. In the following, we work at zero temperature $\beta \rightarrow \infty$.% for simplicity.

{\it Effective action and Ward identity.---} The quench protocol explicitly breaks the time translation symmetry but preserves the space translation symmetry, so unlike the ordinary QED, the Lorentz symmetry is broken. Without Lorentz symmetry, the time and space component couplings can be different in general.
To account for this space and time anisotropy, the effective action is generalized to be
\begin{widetext}
\begin{equation}
\begin{aligned}
	S &= S_0 + S_\text{int}, \\
	S_0 &= \int_0^\infty dt \int_{\bm x} \bar \Psi \left(\ba{cccc} \ii \partial_\mu \gamma^\mu & 0 \\ 0 & \ii \partial_\mu \gamma^\mu \ea \right) \Psi + \int_0^\infty dt \int_{\bm x} A_{c,q}\left(g^{\mu\nu} \partial^2 -\partial^\mu \partial^\nu \right) A_{q,\nu}, \\
	S_\text{int} &= \frac{1}{\sqrt2} \int_0^\infty dt \int_{\bm x} \left( e_c \bar \Psi A_{c,0} \gamma^0 \Psi + g_c \bar \Psi A_{c,i} \gamma^i \Psi + e_q \bar \Psi A_{q,0} \rho^x\gamma^0 \Psi + g_q \bar \Psi A_{q,i} \rho^x\gamma^i \Psi \right),\label{equ:action}
\end{aligned}
\end{equation}
\end{widetext}
where $\int_{\bm x} \equiv \int d^{3}x$, $e_{c,q}$ and $g_{c,q}$ denote the gauge coupling for the time and space component, and $\rho^x$ is the Pauli matrix acting on the Keldysh space. The broken Lorentz symmetry allows $e_{c,q}$ and $g_{c,q}$ to flow 
differently in the RG flow.
We also allow the light velocity to flow, which amounts to rescaling $ \partial_i \rightarrow c \partial_i$.

An important consequence of gauge symmetry is the Ward identity. 
In the quench dynamics, special attention should be paid to the boundary at time $t=0$. 
Under the transformation for arbitrary spacetime dependent variable $\theta(x)$, i.e.,
$\Psi \rightarrow e^{\ii \theta} \Psi$,
the theory is invariant. 
This leads to the Ward identity,
\begin{equation}
\begin{aligned}
    &\delta(t-t_1) \ii G(t_1, t_2; \bm k + \bm p) - \delta(t-t_2) \ii G(t_1, t_2; \bm k)\\
    &- \ii c p_i \ii G(t_1 ,t ; \bm k) \gamma^i \ii G(t, t_2; \bm k + \bm p)\\
    &+ \partial_t \ii G(t_1 ,t ; \bm k) \gamma^0 \ii G(t, t_2; \bm k + \bm p) = 0.\label{equ:ward}
\end{aligned}
\end{equation}
%More conveniently, we can express the Ward identity as the following equation,
%\bea
%	\partial_i G_K(t_1, t_2; \bm k) = -\int_0^\infty dt c [G_R(t_1, t; \bm k) \gamma_i G_K(t, t_2; \bm k) +G_K(t_1, t; \bm k) \gamma_i G_A(t, t_2; \bm k) ] \\
%	+ \ii  [ G_R(t_1, 0; \bm k) \gamma_0 \partial_i G_K(0, t_2; \bm k) + G_K(t_1, 0; \bm k) \gamma_0 \partial_i G_A(0, t_2; \bm k)], \\
%	(t_1 - t_2) G_K(t_1, t_2; \bm k) = \ii \int_0^\infty dt [ G_R(t_1, t; \bm k) \gamma^0 G_K(t, t_2; \bm k) + G_K(t_1, t; \bm k) \gamma^0 G_A(t, t_2; \bm k)],
%\eea
%where the second line resulted from the boundary at $t = 0$. The Ward identity that evolves retarded and advanced propagator is identical to the equilibrium case. It is not hard to verify the propagators in the previous section satisfy the above Ward identity.
Integrate time $t_2$ in Eq.~\ref{equ:ward}, and it is direct to conclude that our Ward identity Eq.~\ref{equ:ward} differs from the equilibrium one by a boundary term at $t=0$: $\ii G(t_1 ,0 ; \bm k) \gamma^0 \ii G(0, t_2; \bm k + \bm p)$. 

\begin{comment}
Before proceeding to the renormalization, we need first to check the renormalizability of the theory. 
Ward identities involving 0 electronic external lines and 2 or 4 gauge boson lines ensure the QED is a renormalizable theory. 
Within the Ward identities involving 0 electronic external lines and 2 or 4-gauge boson lines, the UV divergences of the loop diagrams contributing to the 2-boson and 4-boson amplitudes cancel each other, thus ensuring the renormalizability of the theory. Since the quench is only performed on the fermionic degrees of freedom while leaving the gauge bosons unperturbed, one can conclude that the quenched theory is still renormalizable. 
\end{comment}

{\it One-loop calculation.---}
We calculate the Feynman diagrams up to the one-loop order to obtain the RG equation. 
For the interacting part $S_{\text{int}}$ in Eq.~\ref{equ:action}, we expand it up to the third order to obtain the one-loop correction $\delta S$,
\begin{equation}
\delta S=\left\langle S_{\text {int}}\right\rangle_{>}+\frac{\ii}{2}\left\langle S_{\text {int}}^2\right\rangle_{>}-\frac{1}{6}\left\langle S_{\text {int}}^3\right\rangle_{>},
\end{equation}
where $\langle A\rangle_{>}=\int D \phi_{>} D \bar{\Psi}_{>} D \Psi_{>} A e^{i S_0}$ denotes the integration over the fast modes. 
After performing the integration, which is detailed in the Appendix, we arrive at
\begin{widetext}
\begin{equation}
    \begin{aligned}
	\delta S &= \int_0^\infty dt \int_{\bm x} \Big\{ \ii \bar \Psi \left[ \left( \frac{3g_c g_q - e_c e_q}{32 \pi^2 c^3 \sqrt{1 + \Omega^2}} 	+ \frac{3 g_c^2 - e_c^2}{32\pi^2 c^3} \right) \log \Lambda   \partial_t \gamma^0 + \left( \frac{(e_c e_q + g_c g_q)(3+5\Omega^2) }{96\pi^2 c^3 (1 + \Omega^2)^{\frac{3}{2}}} + \frac{e_c^2 + g_c^2 }{32\pi^2 c^3 }\right) \log \Lambda c \partial_i \gamma^i  \right] \Psi  \\
	&+  \left( \frac{e_c e_q}{6\pi^2 c^3 \sqrt{1+ \Omega^2}} \log \Lambda A_{q,0} \left(- c^2 \nabla^2 \right) A_{c,0} + \frac{g_c g_q}{6\pi^2 c^3 \sqrt{1+ \Omega^2}} \log \Lambda A_{q,i} \left(\partial^2 g^{ij}- c^2 \partial^i \partial^j \right) A_{c,j}  \right)  \\
    &+  \left( \frac{-e_c e_q + 3g_c g_q }{16\pi^2 c^3\sqrt{1 + \Omega^2}} 	+ \frac{- e_c^2 + 3g_c^2}{16\pi^2c^3} \right) \log \Lambda  \left( \frac{e_c}{2\sqrt2} \bar \Psi A_{c,0} \gamma^0 \Psi\right)+ \left( \frac{-e_c e_q + 3g_c g_q }{16\pi^2 c^3\sqrt{1 + \Omega^2}} 	+ \frac{- e_c^2 + 3g_c^2}{16\pi^2c^3} \right) \log \Lambda  \left(\frac{e_q}{2\sqrt2} \bar \Psi A_{q,0} \rho^x \gamma^0 \Psi\right)\\
	&+  \left( \frac{7 (e_c e_q + g_c g_q) }{48\pi^2 c^3\sqrt{1 + \Omega^2}} 	+ \frac{ e_c^2 + g_c^2}{16\pi^2c^3} \right) \log \Lambda  \left( \frac{g_c}{2\sqrt2} \bar \Psi A_{c,i} \gamma^i \Psi\right) \left( \frac{7 (e_c e_q + g_c g_q) }{48\pi^2 c^3\sqrt{1 + \Omega^2}} 	+ \frac{ (e_c^2 + g_c^2)(5+\Omega^2)}{48\pi^2c^3 (1+\Omega^2)} \right) \log \Lambda  \left(\frac{g_q}{2\sqrt2} \bar \Psi A_{q,i} \rho^x \gamma^i \Psi\right) \Big\}
     \end{aligned},
\end{equation}
\end{widetext}
where $\Lambda$ is the energy cutoff and $\Omega \equiv \frac{m}{\Lambda}$ is the quench parameter. 
The fact that the vertex renormalization is the same as the fermion self-energy for the time component is due to the Ward identity. 
However, the spatial component is different due to the boundary term in the Ward identity which is discussed in the previous section. 
To make the RG equation manifest, one can define the following effective interaction strength
\bea
	e_1 = e_c^2 \Lambda^{-\epsilon}, \quad e_2 = \frac{e_c e_q}{\sqrt{1+ \Omega^2} } \Lambda^{-\epsilon}, \\
	g_1 = g_c^2 \Lambda^{-\epsilon}, \quad g_2 = \frac{g_c g_q}{\sqrt{1+ \Omega^2} } \Lambda^{-\epsilon}, 
\eea
where $\epsilon = 4-d$, and consequently, the RG equation reads
\bea
	\frac{d \Omega}{dl} &=& \Omega,\label{equ:rg1} \\
	\frac{d c}{dl} &=& \frac{e_1 - g_1+ e_2 -g_2}{16\pi^2 c^3} + \frac{(e_2+g_2)\Omega^2}{48\pi^2c^3(1+\Omega^2)}, \label{equ:rg2}\\
	\frac{d e_1}{d l} &=& \left( \epsilon  - \frac{e_2}{6\pi^2 c^3} \right) e_1, \\
	\frac{d e_2}{d l} &=& \left( \epsilon - \frac{\Omega^2}{1+ \Omega^2} - \frac{e_2}{6\pi^2 c^3}\right) e_2, \\
	\frac{d g_1}{dl} &=& \left( \epsilon + \frac{3 (e_1 - g_1) + 5(e_2 - g_2)}{24\pi^2 c^3} \right) g_1, \\
	\frac{d g_2}{dl} &=& \left( \epsilon -\frac{\Omega^2}{1+ \Omega^2} + f(l) \right) g_2, \label{equ:rg6}
\eea
where $ f(l)=\frac{11e_1+20e_2-13g_1-20g_2+8(e_1+g_1)/(1+\Omega^2)}{96\pi^2 c^3}$. Moreover, the anomalous dimension is 
\bea
    && \eta_f = \frac{3g_2 - e_2}{32 \pi^2 c^3} + \frac{3g_1 - e_1}{32 \pi^2 c^3}, \\
    && \eta_{b,t} = \frac{e_2}{6\pi^2 c^3}, \quad \eta_{b,s} = \frac{g_2}{6\pi^2 c^3}.
\eea

\begin{figure}
	\includegraphics[width=6cm]{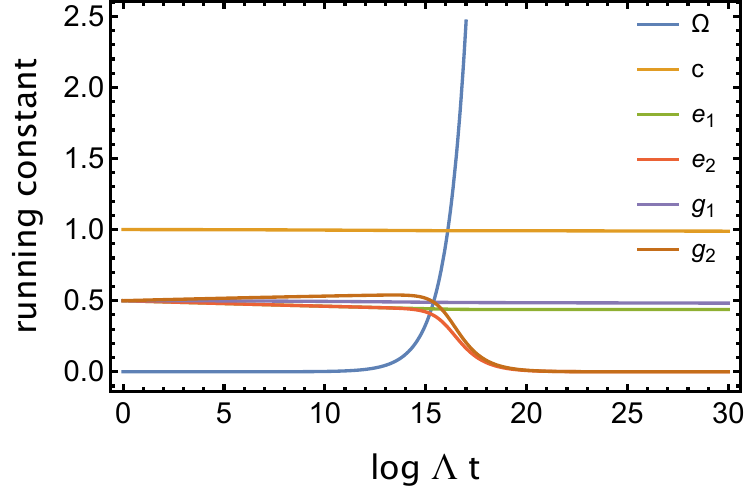}
\caption{Running coupling constants at four dimensions. A prethermalization regime exists controlled by initial interaction values and the quench parameter $m$. After that, there is a new nonthermal fixed point. 	\label{fig:4d}} 
\end{figure}

We focus on $d=4$ ($\epsilon=0$) where the interacting terms are marginally irrelevant in equilibrium. 
The full numerical solution of the $d=4$ coupled RG equation is shown in~Fig.~\ref{fig:4d}. As shown in Fig.~\ref{fig:4d}, the relevant quench parameter $\Omega(t) = \Omega_0 \Lambda t $ controls the renormalization group flow and separates the RG flow into two regimes. 

At short time scales $\Lambda t < \frac1{\Omega_0}$, to analyze the prethermal behavior, we first solve the prethermal fixed point with $\Omega=0$, and find that only a trivial isotropic fixed point with $g_c=g_q=e_q=e_c=0$ exists, and this is exactly the regular QED fixed point in the 3+1 dimension. 
However, as shown in Fig.~\ref{fig:4d}, the interactions remain near the initial value instead of flowing to the fixed point. 
The reason is that in the short timescale, $\Omega$ is small and  $l=\log \Lambda t \sim -\log \Omega_0 \ll \frac{6\pi^2c^3}{e_0}$, thus this trivial fixed point will never be reached in this prethermal timescale.

At the low-energy scale $l \rightarrow \infty$, equivalently, the late timescale, a nontrivial nonthermal fixed point exists. 
We can see from Fig.~\ref{fig:4d} that a transient behavior occurs when the effective quench parameter starts to grow.
After that, the couplings stay at the nonthermal fixed point.
To obtain this fixed point, we solve the RG equations Eq.~\ref{equ:rg1} to Eq.~\ref{equ:rg6}. 
According to the numerical results shown in Fig.~\ref{fig:4d}, we assume a constant light velocity $c$. 
As detailed in the Appendix, the nonthermal fixed point is given by
\begin{equation}
    \begin{aligned}
        g_1^\ast=e_1^\ast &= \frac{e_{1}(0)}{1+ e_2(0) \frac{ \sqrt{1+\Omega_0^2}}{12\pi^2c^3} \log\left( \frac{\sqrt{1+\Omega_0^2}+1}{\sqrt{1+\Omega_0^2}-1} \right)}, \\
  g_2^\ast= e_2^\ast &= 0.\label{equ:nonthermal}
    \end{aligned}
\end{equation}

This result also justifies that the light velocity goes to a constant, as one can check that the right-hand side of Eq.~\ref{equ:rg2} vanishes at a fixed point. 
Therefore, we arrive at a self-consistent nonthermal fixed point. 
At long timescale, only the classical Keldysh interacting term survives, and the fermion anomalous dimension at the fixed point is given by $\eta_f=\frac{3g_1-e_1}{32\pi^2c^3}+\frac{3g_2-e_2}{32\pi^2c^3}=\frac{e^*_1}{16\pi^2c^3}$ where $e_1^*$ is given in Eq.~\ref{equ:nonthermal}. 
The non-trivial anomalous dimension varies continuously as the initial value of the interaction and the quench parameter $m$ vary. 
The information of the initial quench is preserved in the postquench fermion anomalous dimension, and the system exhibits a memory effect. 
Besides, a finite anomalous dimension led to the absence of a quasiparticle pole, which is the signature of non-Fermi liquid behaviors.

Such an interacting nonthermal fixed point can be observed in experiments.
Near the nonthermal fixed point, the information of the initial quench is preserved in the fermion anomalous dimension,
We consider the single-particle correlation function at this fixed point to be approximately given by $G(p) \sim \frac{p \cdot \gamma}{(p^2)^{1-\eta_f}}$, $p=(\omega, \vec p)$. 
After integrating the spectral function over the momentum, it leads to a non-analytical local density of state
\begin{equation}
    \rho(\epsilon) \propto |\epsilon|^{2+2\eta_f},
\end{equation} 
where $\epsilon$ denotes the energy scale. 
As a result, the nonthermal fixed point can be detected from the differential conductance $\frac{dI}{dV} \sim V^{2+2\eta_f} $ in, for example, a scanning tunneling microscope.
Here, $I$ and $V$ denote the current and voltage, respectively.

{\it Discussions.---} 
To summarize, we explore nonequilibrium quantum electrodynamics following a global quench, revealing a noteworthy quantum memory effect in the form of a critical exponent, i.e., the fermion anomalous dimension. 
While our discussion is mainly focused on the 3+1 dimensional case, we conclude this paper by highlighting a few remaining open questions, including the 2+1 dimensional case. 
In equilibrium, quantum electrodynamics in 2+1 dimensions exhibits a nontrivial fixed point. 
We initiate an investigation of quench dynamics in 2+1 dimensions in the Appendix, where the quench trivializes the interacting fixed point, and the validity of the RG flow needs further investigation. 
Conducting a higher-order loop calculation and taking the large-N limit could provide valuable insights into the fate of the nonequilibrium fixed point in 2+1 dimensions. 
We leave such investigation to future work.
Furthermore, it would be of immense significance to experimentally implement the quench protocol examined in this study. 
Potential experimental platforms include programmable cold atom systems and pump-probe experiments in condensed matter systems.

{\it Acknowledgements.---} The work of MRL is supported in part by the NSFC under Grant No. 11825404. % (MRL).
%, the MOSTC Grants No. 2018YFA0305604 and No. 2021YFA1400100 (HY) the CAS Strategic Priority Research Program under Grant No. XDB28000000 (HY). 
MRL acknowledges the support from the Lavin-Bernick Grant during his visit to Tulane university.
The work of SKJ is supported by a startup fund at Tulane University.

\onecolumngrid

\clearpage
\widetext

\begin{center}
\textbf{\large The Supplemental Materials for “Quantum electrodynamics under a quench”}
\end{center}

\renewcommand{\thefigure}{S\arabic{figure}}
\setcounter{figure}{0}
\renewcommand{\theequation}{S\arabic{equation}}
\setcounter{equation}{0}
\renewcommand{\thesection}{\Roman{section}}
\setcounter{section}{0}
\setcounter{secnumdepth}{4}

% \tableofcontents
% \hypersetup{linkcolor=blue}

\subsection{Propagators in Keldysh countour}
\begin{figure}[b]
	\includegraphics[width=6cm]{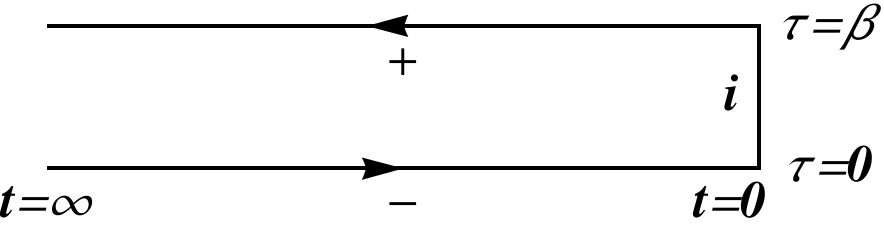} 
	\caption{A schematic contour of the quench dynamics. 
    The vertical (horizontal) line denotes the imaginary-time (real-time Keldysh) evolution.\label{fig:contoursup}}
\end{figure}
To study the quench dynamics of the Dirac fermion, we adopt the Keldysh contour as shown in Fig.~\ref{fig:contoursup}, where the vertical line indicates the generation of the prequench state  via imaginary time evolution. 
The dynamics of the system is determined by the action:
\begin{equation}
\begin{aligned}
S_E & =\int_0^\beta d \tau L_E\left[\bm{A}, \psi_i^{\dagger}, \psi_i\right], \\
S_R & =\int_0^{\infty} d t\left(L\left[\bm{A}_{+}, \psi_{+}^{\dagger}, \psi_{+}\right]-L\left[\bm{A}_{-}, \psi_{-}^{\dagger}, \psi_{-}\right]\right),
\end{aligned}
\end{equation}
where `E' and `R' denote the Euclidean and real-time action separately. The Lagrangians are

\bea
L_E\left[\phi, \psi^{\dagger}, \psi\right] & =&L_{E, b}[\phi]+L_{E, f}\left[\psi^{\dagger}, \psi\right], \\
L_{E, b}[\phi] & =&\frac{1}{4} \int_x F_{\mu \nu} F^{\mu \nu}=-\frac{1}{2} \int_x A_\mu\left(\partial^2 g^{\mu \nu}-\partial^\mu \partial^\nu\right) A_\nu=\frac{1}{2} \int_x A_\mu k^2 g^{\mu \nu} A_\nu, \\
L_{E, f}\left[\psi^{\dagger}, \psi\right] & =&\int_x \psi^{\dagger}\left(\partial_\tau+H_E\right) \psi, \\
L\left[\phi, \psi^{\dagger}, \psi\right] & =&\int_x\left(\frac{1}{2}A_\mu k^2 g^{\mu \nu} A_\nu+\psi^{\dagger}\left(i \partial_t-H_R\right) \psi+e\overline{\Psi}A_\mu \gamma^\mu \Psi\right),
\eea
where the Hamiltonians of the fermion in imaginary-time  and real-time are
\bea \label{eq:dirac_free_hamiltoniansup}
	H_{E} =  \bm{k} \cdot \bm{\alpha} + m \gamma^0, \quad H_{R} =  \bm{k} \cdot \bm{ \alpha}.
\eea
The Dirac matrices satisfies $\{ \gamma^\mu, \gamma^\nu\} = 2  g^{\mu\nu}$ with the convention $g_{\mu\nu} = \text{diag}(1,-1,-1,-1)$. ${\bm \alpha}$ is defined via ${\bm \alpha} \equiv (\gamma^0 \gamma^1, \gamma^0 \gamma^2, \gamma^0 \gamma^3)$. 
Applying the Keldysh rotation
\begin{equation}
	\psi_{c/q} = \frac1{\sqrt2} (\psi_+ \pm \psi_-), \quad \psi_{c/q}^\dag = \frac1{\sqrt2} (\psi_+^\dag \mp \psi_-^\dag),\quad A_{c/q} = \frac1{\sqrt2} (A_+ \pm A_-),
\end{equation}
the real-time action is reduced to
\begin{equation}
\begin{aligned}
	S &= S_0 + S_\text{int}, \\
	S_0 &= \int_0^\infty dt \int_{\bm x} \bar \Psi \left(\ba{cccc} \ii \partial_\mu \gamma^\mu & 0 \\ 0 & \ii \partial_\mu \gamma^\mu \ea \right) \Psi + \int_0^\infty dt \int_{\bm x} A_{c,q}\left(g^{\mu\nu} \partial^2 -\partial^\mu \partial^\nu \right) A_{q,\nu}, \\
	S_\text{int} &= \frac{1}{\sqrt2} \int_0^\infty dt \int_{\bm x} \left( e_c \bar \Psi A_{c,0} \gamma^0 \Psi + g_c \bar \Psi A_{c,i} \gamma^i \Psi + e_q \bar \Psi A_{q,0} \rho^x\gamma^0 \Psi + g_q \bar \Psi A_{q,i} \rho^x\gamma^i \Psi \right).\label{equ:action1sup}
\end{aligned}
\end{equation}
According to the Keldysh field theory, the propagators are given by 
\begin{equation}
\begin{aligned}
\hat{G} & =-i\left\langle\Psi \Psi^\dagger \right\rangle=\left(\begin{array}{cc}
G_R & G_K \\
0 & G_A
\end{array}\right), \quad \hat{D} & =-i\left\langle A_\mu A_\nu\right\rangle=\left(\begin{array}{cc}
D_K & D_R \\
D_A & 0
\end{array}\right) .\label{equ:fbpropagator}
\end{aligned}
\end{equation}
in the classical/quantum fields basis. 
We introduce the source field $\xi_{c/q}$ that couples to the fermionic field $\psi_{c/q}$
\begin{equation}
W\left[\xi_{c / q}\right]=\ln \int D \mu_f \exp \left[-\int d \tau L_{E, f}\left[\psi_i^{\dagger}, \psi_i\right]+i \int d t\left(L_{0, f}\left[\Psi^{\dagger}, \Psi\right]+\left(\xi_c^{\dagger} \psi_c+\xi_q^{\dagger} \psi_q+H . c .\right)\right)\right],\label{equ:generator}
\end{equation}
where $\int D \mu_f \equiv D\psi_{i}^{\dagger}D\psi_i D\psi_{c}^{\dagger}D\psi_c D\psi_{q}^{\dagger}D\psi_q$. 
By integrating the dynamical fields, one can get the fermion propagators. Integrating over $\psi_i^{\dagger}$, we can obtain the equation of motion of $\psi_i$: $(\partial_{\tau} + \mathcal{H})\psi_i=0$. The wave function can be solved directly. We assume the value of $\psi$ at time $\tau =0$ to be $\psi_0$, then the solution to the equation of motion is
\bea
	\psi(\tau) = \left(e^{- \xi \tau} P_+(m) + e^{\xi \tau} P_-(m) \right) \psi_0,
\eea
where $P_{+/-}(m)$ are projectors that project to the positive/negative energy eigenstate:
\begin{equation}
\begin{aligned}
P_{\pm}(m)&= \frac12 \left( 1 \pm ( \bm{\hat k}_m \cdot \bm{\alpha} + \hat m \gamma^0) \right), \\
H_{E} P_{\pm}(m) &= \pm \xi P_{\pm}(m), \quad H_R  P_{\pm} = \pm |\bm{k}| P_{\pm},
\end{aligned}
\end{equation}
with $\xi=\sqrt{k^2+m^2}$. 
With the boundary condition $\psi_{-}(0)=-\psi_i(0),\psi_+(0)=\psi_i(\beta)$, the initial values of $\psi_{c/q}$ are given by
\bea
	\psi_{c0} &=& \frac1{\sqrt2 } (\psi(\beta) - \psi(0))=\frac1{\sqrt2 } \left(\left(e^{- \xi \beta} P_+(m) + e^{\xi \beta} P_-(m) \right)-1\right)\psi_0,\label{equ:boundaryc} \\
\psi_{q0} &=& \frac1{\sqrt2 } (\psi(\beta) + \psi(0))=\frac1{\sqrt2 } \left(\left(e^{- \xi \beta} P_+(m) + e^{\xi \beta} P_-(m) \right)+1\right)\psi_0,\label{equ:boundaryq}\\
	\psi_{c0}^\dagger &=& \frac1{\sqrt2 } (\psi^\dagger(\beta) + \psi^\dagger(0))=\frac1{\sqrt2 } \psi_0^\dagger\left(\left(e^{- \xi \beta} P_+(m) + e^{\xi \beta} P_-(m) \right)+1\right),\label{equ:boundarycd} \\
\psi_{q0} &=& \frac1{\sqrt2 } (\psi^\dagger(\beta) - \psi^\dagger(0))=\frac1{\sqrt2 } \psi_0^\dagger\left(\left(e^{- \xi \beta} P_+(m) + e^{\xi \beta} P_-(m) \right)-1\right).\label{equ:boundaryqd}
\eea 
After the integration of the fields $\psi_i$ and $\psi_i^\dag$, the Eq.~\ref{equ:generator} is reduced to
\begin{equation}
W\left[\xi_{c / q}\right]=\ln \int D \psi_0 D \psi_c^{\dagger} D \psi_c D \psi_f^{\dagger} D \psi_f \exp \left[i \int d t\left(\int_{\bm x} \bar \Psi \left(\ba{cccc} \ii \partial_\mu \gamma^\mu & 0 \\ 0 & \ii \partial_\mu \gamma^\mu \ea \right) \Psi+\left(\xi_c^{\dagger} \psi_c+\xi_q^{\dagger} \psi_q+\text { H.c. }\right)\right)\right].
\end{equation}
Again we integrate $\psi_c^\dagger$ and $\psi_q$ and obtain the equation of motion of $\psi_c$ and $\psi_q^\dagger$:
\bea
\left(i \partial_t-\mathcal{H}\right) \psi_c+\xi_c=0, \\
-i \partial_t \psi_q^{\dagger}-\psi_q^{\dagger} \mathcal{H}+\xi_q^{\dagger}=0.
\eea
Using the boundary condition Eq.~\ref{equ:boundaryc} and ~\ref{equ:boundaryq}, the solutions can be obtained directly
\begin{equation}
\begin{aligned}
\psi_c(t) & =\left[P_+\left(e^{- \xi \beta} P_+(m) + e^{\xi \beta} P_-(m) -1\right)e^{-ikt}+P_-\left(e^{- \xi \beta} P_+(m) + e^{\xi \beta} P_-(m) -1\right)e^{ikt}\right]\psi_0 -\int d t^{\prime} G_R\left(\bm k, t-t^{\prime}\right) \xi_c\left(t^{\prime}\right), \\
\psi_q^{\dagger}(t) & =\psi_0^{\dagger}\left[\left(e^{- \xi \beta} P_+(m) + e^{\xi \beta} P_-(m) -1\right)P_+e^{ikt}+\left(e^{- \xi \beta} P_+(m) + e^{\xi \beta} P_-(m) -1\right)P_-e^{-ikt}\right]-\int d t^{\prime} \xi_q^{\dagger}\left(t^{\prime}\right) G_A\left(\bm k, t^{\prime}-t\right),\label{equ:solutionforpsicq}
\end{aligned}
\end{equation}
where $P_{\pm}\equiv P_{\pm}(0)$, and $G_{R/A}$ are the retarded and advanced Green's function given by
\begin{equation}
\begin{aligned}
& G_R\left(\bm{k}, t-t^{\prime}\right)=-i \Theta\left(t-t^{\prime}\right)\left(e^{-i |k|\left(t-t^{\prime}\right)} P_{+}(k)+e^{i |k|\left(t-t^{\prime}\right)} P_{-}(k)\right), \\
& G_A\left(\bm{k}, t-t^{\prime}\right)=i \Theta\left(t^{\prime}-t\right)\left(e^{-i |k|\left(t-t^{\prime}\right)} P_{+}(k)+e^{i |k|\left(t-t^{\prime}\right)} P_{-}(k)\right).
\end{aligned}
\end{equation}
Substituting the solution Eq.~\ref{equ:solutionforpsicq} into the action and integrating over $\psi_c$ and $\psi_{q}^\dagger$, we arrive at

\begin{equation}
\begin{aligned}
e^{W\left[\xi_{c / q}\right]}= & \exp -i \int d t d t^{\prime} \int_k\left[\xi_c^{\dagger}(t) G_R\left(k, t-t^{\prime}\right) \xi_c\left(t^{\prime}\right)+\xi_q^{\dagger}(t) G_A\left(k, t-t^{\prime}\right) \xi_q(t)\right] \\
& \times \int D \psi_0 \exp \left(\psi_{0 q}^{\dagger} \psi_{0 q}\right) \exp \frac{i}{\sqrt{2}} \int d t \int_k\left[\xi_c^{\dagger}[P_+(e^{- \xi \beta} P_+(m) + e^{\xi \beta} P_-(m) -1\right)e^{-ikt}\\
&+P_-\left(e^{- \xi \beta} P_+(m) + e^{\xi \beta} P_-(m) -1\right)e^{ikt}]\psi_0+ \\
& \psi_0^{\dagger}\left[\left(e^{- \xi \beta} P_+(m) + e^{\xi \beta} P_-(m) -1\right)P_+e^{ikt}+\left(e^{- \xi \beta} P_+(m) + e^{\xi \beta} P_-(m) -1\right)P_-e^{-ikt}\right] \xi_q].
\end{aligned}
\end{equation}
%within the boundary condition Eq.~\ref{equ:boundaryq}, 
Further integrating over $\psi_0$ leads to the generating function for free propagators:
\begin{equation}
W\left[\xi_{c / q}\right]= \int d t d t^{\prime} \int_k\left(-i\left[\xi_c^{\dagger}(t) G_R\left(k, t-t^{\prime}\right) \xi_c\left(t^{\prime}\right)+\xi_q^{\dagger}(t) G_A\left(k, t-t^{\prime}\right) \xi_q(t)\right]-\xi_c^{\dagger}(t) G_K(k,t-t') \xi_q\left(t^{\prime}\right)\right),
\end{equation}
where
\begin{equation}
\begin{aligned}
	\ii G_K(t_1, t_2; \bm{k}) = \tanh \frac{\beta \xi}2(e^{-\ii |\bm{k}|t_1} P_+ + e^{\ii |\bm{k}|t_1} P_-)
(P_+(m) - P_-(m)) (e^{\ii |\bm{k}|t_2} P_+ + e^{-\ii |\bm{k}|t_2} P_-). %\\
%	= \tanh \frac{\beta \xi}2 \left(\cos \chi (e^{-\ii |\bm{k}|(t_1-t_2)} P_+ - e^{\ii |\bm{k}|(t_1-t_2)} P_-)\gamma^0 + \sin \chi (e^{\ii |\bm{k}|(t_1+t_2)} P_+ + e^{-\ii |\bm{k}|(t_1+t_2)} P_-) \right)%
\end{aligned}
\end{equation}

Next, we turn to the boson propagator. The action of the free gauge boson is 
\bea
L_{E, b}[\phi] & =&\frac{1}{4} \int_x F_{\mu \nu} F^{\mu \nu}=-\frac{1}{2} \int_x A_\mu\left(\partial^2 g^{\mu \nu}-\partial^\mu \partial^\nu\right) A_\nu=\frac{1}{2} \int_x A_\mu k^2 g^{\mu \nu} A_\nu,
\eea
The retarded Green's function of the gauge field is given by
\bea
D_{\mu \nu}^R(\omega,\bm k)=-g_{\mu \nu} \frac{1}{(\omega+i \delta-k)(\omega+i \delta+k)}=-g_{\mu \nu} \frac{1}{2 k}\left(\frac{1}{\omega+i \delta-k}-\frac{1}{\omega+i \delta+k}\right).
\eea
One can obtain the real-time retarded propagator by Fourier transformation
\begin{equation}
    D_{\mu \nu}^R(t,\bm k)  =-g_{\mu \nu} \frac{1}{2 k} \int\left(\frac{1}{\omega+i \delta-k}-\frac{1}{\omega+i \delta+k}\right) e^{-i \omega t}=-g_{\mu \nu} \theta[t]\left(-\frac{\operatorname{Sin}[k t]}{k}\right).
\end{equation}
The same calculation also follows for the advanced propagator
\begin{equation}
\begin{aligned}
&D_{\mu \nu}^A(\omega,\bm k)=-g_{\mu \nu} \frac{1}{(\omega-i \delta-k)(\omega-i \delta+k)}=-g_{\mu \nu} \frac{1}{2 k}\left(\frac{1}{\omega-i \delta-k}-\frac{1}{\omega-i \delta+k}\right)\\
&D_{\mu \nu}^A(t,\bm k)=-g_{\mu \nu} \frac{1}{2 k} \int\left(\frac{1}{\omega-i \delta-k}-\frac{1}{\omega-i \delta+k}\right) e^{-i \omega t}=-g_{\mu \nu} \theta[-t] \frac{\sin [k t]}{k}
\end{aligned}
\end{equation}
Thus, we can write the time-ordered propagator
\begin{equation}
    D_{\mu\nu}^t(t,\bm k)=-g_{\mu\nu}\frac{-\ii}{2k}(\theta[t]e^{-\ii kt}+\theta[-t]e^{\ii kt}),
\end{equation}
and the two-point functions are
\begin{equation}
    \left\langle A_\mu(t) A_v(0)\right\rangle=-g_{\mu v} \frac{e^{-i k t}}{2 k},\left\langle A_v(0) A_\mu(t)\right\rangle=-g_{\mu v} \frac{e^{i k t}}{2 k}.
\end{equation}
According to Eq.~\ref{equ:fbpropagator}, the Keldysh propagator is 
\begin{equation}
\begin{aligned}
i D_{\mu v}^K&=\left\langle A_c(t) A_c(0)\right\rangle=\frac{1}{2}\left\langle\left(A_{+}(t)+A_{-}(t)\right)\left(A_{+}(0)+A_{-}(0)\right)\right\rangle =\langle A(t) A(0)+A(0) A(t)\rangle -g_{\mu v} \frac{\operatorname{cos}[k t]}{k}.
\end{aligned}
\end{equation}

For later convenience, we modify the definition of the fermion propagators by attaching a $\gamma_0$ matrix to the original definition:
\bea
\hat{G} & =-i\left\langle\Psi \overline{\Psi} \right\rangle=\left(\begin{array}{cc}
G_R & G_K \\
0 & G_A
\end{array}\right),
\eea
with
\bea
& G_R\left(\bm{k}, t-t^{\prime}\right)=-i \Theta\left(t-t^{\prime}\right)\left(e^{-i |k|\left(t-t^{\prime}\right)} P_{+}(k)+e^{i |k|\left(t-t^{\prime}\right)} P_{-}(k)\right)\gamma^0, \\
& G_A\left(\bm{k}, t-t^{\prime}\right)=i \Theta\left(t^{\prime}-t\right)\left(e^{-i |k|\left(t-t^{\prime}\right)} P_{+}(k)+e^{i |k|\left(t-t^{\prime}\right)} P_{-}(k)\right)\gamma^0,\\
&G_K(t_1, t_2; \bm{k}) = -\ii\tanh \frac{\beta \xi}2(e^{-\ii |\bm{k}|t_1} P_+ + e^{\ii |\bm{k}|t_1} P_-)
(P_+(m) - P_-(m)) (e^{\ii |\bm{k}|t_2} P_+ + e^{-\ii |\bm{k}|t_2} P_-) \gamma^0.
\eea

\subsection{\label{app:ward}Ward identity}

In this section, we derive the Ward identity for the quenched fermion system. We start with the fermion action in equilibrium $S=\int\overline{\psi}\partial_\mu\gamma^\mu\psi$. 
The equilibrium fermion propagator is given by $G^{-1}=\ii\partial_\mu \gamma^{\mu}=k_\mu \gamma^{\mu}$, $ G=\frac{k_\mu\gamma^\mu}{\bm{k}^2}$, within which the perturbative Ward identity can be derived:
\begin{equation}
\begin{aligned}
    G(\bm{k+p})-G(\bm{k})&=\frac{(k_\mu+p_{\mu})\gamma^\mu}{(\bm{k}+\bm{p})^2}-\frac{k_\mu\gamma^\mu}{\bm{k}^2}=\frac{\bm{k}^2(k_\mu+p_\mu)\gamma^{\mu}-(\bm{k}+\bm{p})^2k_\mu\gamma^\mu}{(\bm{k}+\bm{p})^2\bm{k}^2}\\
    &=\frac{\bm{k}^2p_\mu\gamma^\mu -2\bm{k}\cdot\bm{p}k_\mu\gamma^\mu-\bm{p}^2k_\mu\gamma^\mu}{(\bm{k}+\bm{p})^2\bm{k}^2}=-\frac{k_\nu\gamma^\nu p_\mu\gamma^\mu k_\rho\gamma^\rho+k_\mu\gamma^\mu p^2}{(\bm{k}+\bm{p})^2\bm{k}^2}\\
    &=-\frac{k_\nu\gamma^\nu p_\mu\gamma^\mu (k_\rho\gamma^\rho+p_\rho\gamma^\rho)}{(\bm{k}+\bm{p})^2\bm{k}^2}=-p_\mu \frac{k_\nu \gamma^\nu}{\bm{k}^2}\gamma^\mu\frac{(k_\rho+p_\rho)\gamma^\rho}{(\bm{k}+\bm{p})^2}\\
    &=-p_\mu G(\bm{k})\gamma^\mu G(\bm{k}+\bm{p}).
\end{aligned}
\end{equation}
Thus, one can obtain the Ward identity in the frequency-momentum space $\partial^\mu G(\bm{k})=-G(\bm{k})\gamma^\mu G(\bm{k})$. 
However, in our case, the quench breaks the time translation symmetry explicitly, so we turn to the real-time coordinate
\begin{equation}
\begin{aligned}
    \int dt_2G(t_1-t_2,\bm{k})\gamma^\mu G(t_2-t_3,\bm{k})&=\int\frac{d\omega_1d\omega_2}{4\pi^2}\int dt_2 \exp{[-\ii\omega_1(t_1-t_2)-\ii\omega_2(t_2-t_3)]}G(\omega_1,\bm{k})\gamma^\mu G(\omega_2,\bm{k})\\
    &=\int\frac{d\omega}{2\pi}G(\omega,\bm{k})\gamma^\mu G(\omega,\bm{k})e^{-\ii\omega(t_1-t_3)}=-\int\frac{d\omega}{2\pi}\partial_{k_\mu} G(\omega,\bm{k})e^{-\ii\omega(t_1-t_3)}\\
    &=-\partial_{k_\mu} G(t_1-t_3,\bm{k}).
\end{aligned}
\end{equation}
In particular, we can write the above equation in another form
\begin{equation}
\begin{aligned}
    G(t_1, t_3; \bm k + \bm p) - G(t_1, t_3; \bm k)= -\int dt_2 \left[c p_i G(t_1 ,t_2 ; \bm k) \gamma^i G(t_2, t_3; \bm k + \bm p)+\ii \partial_{t_2} G(t_1 ,t_2 ; \bm k) \gamma^0 G(t_2, t_3; \bm k + \bm p)\right].\label{equ:wardsup}
\end{aligned}
\end{equation}
where the second term of the r.h.s is the surface term and will disappear in equilibrium. 
% The
% \begin{equation}
%     \delta(t-t_1) \ii G(t_1, t_2; \bm k + \bm p) - \delta(t-t_2) \ii G(t_1, t_2; \bm k)- \ii c p_i \ii G(t_1 ,t ; \bm k) \gamma^i \ii G(t, t_2; \bm k + \bm p)+ \partial_t \ii G(t_1 ,t ; \bm k) \gamma^0 \ii G(t, t_2; \bm k + \bm p) = 0.\label{equ:ward}
% \end{equation}

\subsection{\label{app:rg}RG analysis}

In this section, we give a brief derivation of the RG equations. As derived in the main text, the Keldysh action for the fermion coupled to gauge boson is given by
\begin{equation}
\begin{aligned}
	S &= S_0 + S_\text{int}, \\
	S_0 &= \int_0^\infty dt \int_{\bm x} \bar \Psi \left(\ba{cccc} \ii \partial_\mu \gamma^\mu & 0 \\ 0 & \ii \partial_\mu \gamma^\mu \ea \right) \Psi + \int_0^\infty dt \int_{\bm x} A_{c,q}\left(g^{\mu\nu} \partial^2 -\partial^\mu \partial^\nu \right) A_{q,\nu}, \\
	S_\text{int} &= \frac{1}{\sqrt2} \int_0^\infty dt \int_{\bm x} \left( e_c \bar \Psi A_{c,0} \gamma^0 \Psi + g_c \bar \Psi A_{c,i} \gamma^i \Psi + e_q \bar \Psi A_{q,0} \rho^x\gamma^0 \Psi + g_q \bar \Psi A_{q,i} \rho^x\gamma^i \Psi \right).\label{equ:actionsup}
\end{aligned}
\end{equation}

\begin{figure}[h]
	\includegraphics[width=15cm]{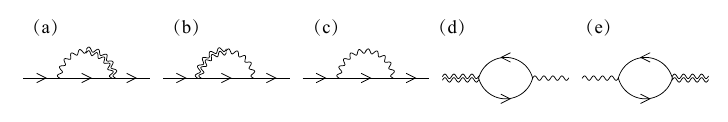} 
	\caption{The Feynman diagrams that correct bosonic and fermionic two-point correlations.\label{fig:2orderdim}}
\end{figure}

We then treat the interaction terms $S_{\text{int}}$ as a perturbation. 
The second-order perturbation gives the correction to the two-point correlations. 
The corresponding Feynman diagrams are shown in Fig.~\ref{fig:2orderdim}. (a), (b), and (c) of Fig.~\ref{fig:2orderdim} are the corrections to the two-point fermion correlator:
\begin{equation}
\begin{aligned}
\text{(a)+(b)+(c)}&  = \frac{i}{2} \frac{e_c e_q}{2} \int \bar{\psi}(t_1)\left\langle\rho^x \gamma^\mu \psi(t_1) \bar{\psi}(t_2) \gamma^\nu A_{q, \mu}(t_1) A_{c, \nu}(t_2)\right\rangle \psi(t_2)\\
&+\frac{i}{2} \frac{e_c e_q}{2} \int \bar{\psi}(t_1)\left\langle\gamma^\mu \psi(t_1) \bar{\psi}(t_2) \rho^x \gamma^\nu A_{c, \mu}(t_1) A_{q, v}(t_2)\right\rangle \psi(t_2) \\
&+\frac{i}{2} \frac{e_c^2}{2} \int \bar{\psi}(t_1)\left\langle\gamma^\mu \psi(t_1) \bar{\psi}(t_2) \gamma^\nu A_{c, \mu}(t_1) A_{c, v}(t_2)\right\rangle \psi(t_2)\\
& =\frac{i}{2} \frac{e_c e_q}{2} \int \bar{\psi}(t_1)\frac{1}{4}\operatorname{Tr}\left [\gamma^\mu(i G(t_1, t_2)) \rho^x \gamma^\nu\right ]\left(\dot{i} D_{\mu \nu}^R (t_1, t_2)\right) \psi(t_2)\\
&+\frac{i}{2} \frac{e_c e_q}{2} \int \bar{\psi}(t_1)\frac{1}{4}\operatorname{Tr}\left [\rho^x \gamma^\mu(\dot{i} G(t_1, t_2)) \gamma^\nu\right ]\left(i D_{\nu \mu}^R(t_2, t_1)\right) \psi(t_2)\\
&+\frac{i}{2} e_c^2 \int \bar{\psi}(t_1) \gamma^\alpha \psi(t_2) \frac{1}{4} \operatorname{Tr}\left[\gamma_\alpha \gamma^\mu(i G(t_1, t_2)) \gamma^\nu\right]\left(i D_{\nu \mu}^K(t_2,t_1)\right).\label{equ:ferana}
\end{aligned}
\end{equation}

Substituting the exact forms of the propagators into the Eq.~\ref{equ:ferana}, one can obtain
\begin{equation}
\begin{aligned}
\text{(a)+(b)+(c)}& = i \int \bar{\Psi}\left(\frac{(3 g_c g_q-e_c e_q)}{32 c^3 \sqrt{k^2+m^2} \pi^2} \gamma^0 \partial_t+\frac{\left(3 g_c^2-e_c^2\right)}{32 c^3 k \pi^2} \gamma^0 \partial_t\right) \Psi\\
&+i \int \bar{\Psi} \left(\frac{(e_c e_q+g_c g_q)\left(3 k^2+5 m^2\right)}{96 c^3\left(k^2+m^2\right)^{3 / 2} \pi^2} c \gamma^i \partial_i+\frac{\left(e_c^2+g_c^2\right)}{32 c^3 k \pi^2} c \gamma^i \partial_i\right) \Psi.\label{equ:fermion}
\end{aligned}
\end{equation}
The light velocity $c$ is renormalized by:
\begin{equation}
\frac{dc}{dl}=\left(\frac{(e_ce_q+g_cg_q)\left(3+5 \Omega^2\right)}{96 \pi^2 c^3\left(1+\Omega^2\right)^{3 / 2}}-\frac{3 g_c g_q-e_c e_q}{32 \pi^2 c^3 \sqrt{1+\Omega^2}}+\frac{e_c^2-g_c^2}{16 c^3 \pi^2}\right) c,
\end{equation}
where $\Lambda$ is the energy cutoff and $\Omega \equiv \frac{m}{\Lambda}$ is the quench parameter. Then the fermionic anomalous dimension is $\eta_f=\frac{3g_c g_q-e_c e_q} {32 \pi^2 c^3 \sqrt{1+\Omega^2}}+\frac{3 g_c^2-e_c^2}{32 \pi^2 c^3}$.

In the same way, one can calculate the diagrams (d) and (e) in Fig.~\ref{fig:2orderdim} and obtain the 2-point corrections to the gauge boson:
\begin{equation}
\begin{aligned}
(d) + (e) & =\frac{i}{2}\left\langle\left(\frac{e_c}{\sqrt{2}} \int A_{c, \mu} \bar{\psi} \gamma^\mu \psi+\frac{e_q}{\sqrt{2}} \int A_{q, \mu} \bar{\psi} \rho^x \gamma^\mu \psi\right)^2\right\rangle\\
&= \frac{i}{2} e_c e_q \int A_{q, \mu} A_{c, \nu}\left\langle\bar{\psi} \rho^x \gamma^\mu \psi \bar{\psi} \gamma^v \psi\right\rangle\\
&= 
-\frac{i}{2} e_c e_q \int A_{q, \mu}(t_1) A_{c, \nu}(t_2) \operatorname{Tr}\left[\rho^x \gamma^\mu i G(t_1, t_2) \gamma^v i G(t_2, t_1)\right].\label{equ:2boson}
\end{aligned}
\end{equation}
With the quenched propagators, the Eq.~\ref{equ:2boson} can be calculated:
\begin{equation}
\begin{aligned}
\frac{i}{2}\left\langle s_{\text {int }}^2\right\rangle & =\int A_{q, 0}\left(-c^2 \nabla^2\right) A_{c, 0} \int dk \frac{e_c e_q}{6 \pi^2 c^3\left(k^2+m^2\right)^{1 / 2}}\\
&+\int A_{q, y}\left(\partial_t^2-c^2\left(\partial_x^2+\partial_z^2\right)\right) A_{c, y} \int dk \frac{g_c g_q}{6 \pi^2 c^3\left(k^2+m^2\right)^{1 / 2}}\\
&+\int A_{q, y}c^2\nabla^2 A_{c, y}\int dk \frac{\left(m^4-2 k^2 m^2\right)g_cg_q}{15\pi^2c^3\left(k^2+m^2\right)^{5 / 2}} \ldots\label{equ:anoboson}
\end{aligned}
\end{equation}
The last term can be ignored, then the Eq.~\ref{equ:anoboson} gives the boson's anomalous dimension $\eta_{b_t}=\frac{e_ce_q}{6\pi^2c^3(1+\Omega^2)^{1/2}}$, $\eta_{b_{x/y/z}}=\frac{g_cg_q}{6\pi^2c^3(1+\Omega^2)^{1/2}}$.

\begin{figure}[h]
	\includegraphics[width=15cm]{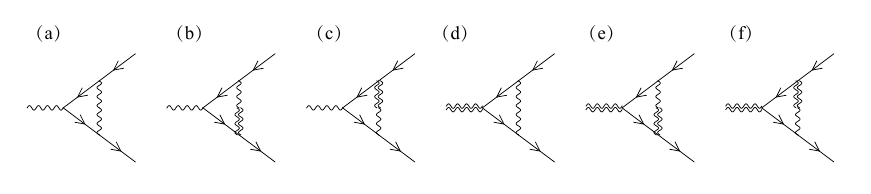} 
	\caption{The Feynman diagrams that correct the interactions.\label{fig:3orderdim}}
\end{figure}

To renormalize the interacting terms $g_{c/q}, e_{c/q}$, one needs to calculate the third-order perturbation of the interactions. The corresponding Feynman diagrams are shown in Fig.~\ref{fig:3orderdim}, and the correction $\delta S_v$ is given by
\begin{equation}
    \begin{aligned}
    \delta S_{v} &= \int_0^\infty dt \int \left( \frac{7 (e_c e_q + g_c g_q) }{48\pi^2 c^3\sqrt{1 + \Omega^2}} 	+ \frac{ e_c^2 + g_c^2}{16\pi^2c^3} \right) \log \Lambda  \left( \frac{g_c}{2\sqrt2} \bar \Psi A_{c,i} \gamma^i \Psi\right).\\
     &+ \int_0^\infty dt \int \left( \frac{7 (e_c e_q + g_c g_q) }{48\pi^2 c^3\sqrt{1 + \Omega^2}} 	+ \frac{ (e_c^2 + g_c^2)(5+\Omega^2)}{48\pi^2c^3(1+\Omega^2)} \right) \log \Lambda  \left(\frac{g_q}{2\sqrt2} \bar \Psi A_{q,i} \rho^x \gamma^i \Psi\right)\nn.
    \end{aligned}
\end{equation}

Notice that the results of the time component $e_c$ and $e_q$ are direct, but one needs to pay additional attention to the space case. According to the Ward identity Eq.~\ref{equ:wardsup}, the vertex contribution of the interacting term differs from the fermionic contribution by a surface contribution. 

With the results obtained above, the flow equations for the interactions can then be derived directly
\begin{equation}
\begin{aligned}
& \frac{de_c}{dl}=\left(\frac{\epsilon}{2}-\frac{\eta_{bt}}{2}-\eta_f\right) e_c+\left(\frac{3 g_c g_q-e_c e_q}{32 \pi^2 c^3 \sqrt{1+\Omega^2}}+\frac{3 g_c^2-e_c^2}{32 \pi^2 c^3}\right) e_c, \\
& \frac{dg_c}{dl}=\left(\frac{\epsilon}{2}-\frac{\eta_{bxyz}}{2}-\eta_f\right) g_c+\left(\frac{7(e_ce_q+g_c g_q)}{96 \pi^2 c^3 \sqrt{1+\Omega^2}}+\frac{e_c^2+g_c^2}{32 \pi^2 c^3}\right) g_c, \\
& \frac{de_q}{dl}=\left(\frac{\epsilon}{2}-\frac{\eta_{bt}}{2}-\eta_f\right) e_c+\left(\frac{3 g_c g_q-e_c e_q}{32 \pi^2 c^3 \sqrt{1+\Omega^2}}+\frac{3 g_c^2-e_c^2}{32 \pi^2 c^3}\right) e_q, \\
& \frac{dg_q}{dl}=\left(\frac{\epsilon}{2}-\frac{\eta_{bxyz}}{2}-\eta_f\right) g_c+\left(\frac{7(e_ce_q+g_c g_q)}{96 \pi^2 c^3 \sqrt{1+\Omega^2}}+\frac{(e_c^2+g_c^2)(5+\Omega^2)}{48 \pi^2 c^3(1+\Omega^2)}\right) g_c. \label{equ:rgs}
\end{aligned}
\end{equation}
To make the RG equations easier to manipulate and get a more direct understanding of the RG flow, one can define
\bea
	e_1 = e_c^2 \Lambda^{-\epsilon}, \quad e_2 = \frac{e_c e_q}{\sqrt{1+ \Omega^2} } \Lambda^{-\epsilon}, \\
	g_1 = g_c^2 \Lambda^{-\epsilon}, \quad g_2 = \frac{g_c g_q}{\sqrt{1+ \Omega^2} } \Lambda^{-\epsilon}, 
\eea
and substitute into Eq.~\ref{equ:rgs}, then the RG equations reduce to
\bea
	\frac{d \Omega}{dl} &=& \Omega,\label{equ:rg1s} \\
	\frac{d c}{dl} &=& \frac{e_1 - g_1+ e_2 -g_2}{16\pi^2 c^3}+ \frac{(e_2+g_2)\Omega^2}{48\pi^2c^3(1+\Omega^2)},\label{equ:rg2s} \\
	\frac{d e_1}{d l} &=& \left( \epsilon  - \frac{e_2}{6\pi^2 c^3} \right) e_1,\label{equ:rg3s} \\
	\frac{d e_2}{d l} &=& \left( \epsilon - \frac{\Omega^2}{1+ \Omega^2} - \frac{e_2}{6\pi^2 c^3}\right) e_2,\label{equ:rg4s} \\
	\frac{d g_1}{dl} &=& \left( \epsilon + \frac{3 (e_1 - g_1) + 5(e_2 - g_2)}{24\pi^2 c^3} \right) g_1, \\
	\frac{d g_2}{dl} &=& \left( \epsilon -\frac{\Omega^2}{1+ \Omega^2} + \frac{11e_1+20e_2-13g_1-20g_2+8(e_1+g_1)/(1+\Omega^2)}{96\pi^2 c^3} \right) g_2.\label{equ:rg6s}
\eea
This leads to the RG equation in the main text.

We are ready to evaluate the non-trivial fixed point presented in the main text.
Assuming a constant light velocity $c$, and notice that Eq.~\ref{equ:rg1s}, Eq.~\ref{equ:rg3s} and Eq.~\ref{equ:rg4s} are decoupled from the rest equations, and one can obtain their solutions for $\epsilon=0$
\bea \label{eq:solution}
	\Omega(l) &=& \Omega_0 e^{l}, \\
	e_1(l) &=& \frac{e_{1}(0)}{1+ e_2(0) \frac{ \sqrt{1+\Omega_0^2}}{12\pi^2 c^3} \log\left( \frac{\sqrt{1+\Omega_0^2}+1}{\sqrt{1+\Omega_0^2}-1} \frac{\sqrt{1+\Omega_0^2 e^{2l}}-1}{\sqrt{1+\Omega_0^2 e^{2l}}+1} \right)} , \\
	e_2(l) &=& \frac{\left( \frac{1+\Omega_0^2}{1+ \Omega_0^2 e^{2l}} \right)^{1/2}e_{2}(0)}{1 + e_2(0) \frac{\sqrt{1+ \Omega_0^2}}{12\pi^2 c^3}  \log\left( \frac{\sqrt{1+\Omega_0^2}+1}{\sqrt{1+\Omega_0^2}-1} \frac{\sqrt{1+\Omega_0^2 e^{2l}}-1}{\sqrt{1+\Omega_0^2 e^{2l}}+1} \right) }.
\eea
Take $l$ tends to infinity, $e_1$ and $e_2$ reaches the fixed point
\begin{equation}
    e_1^\ast=\frac{e_{1}(0)}{1+ e_2(0) \frac{ \sqrt{1+\Omega_0^2}}{12\pi^2c^3} \log\left( \frac{\sqrt{1+\Omega_0^2}+1}{\sqrt{1+\Omega_0^2}-1} \right)}, \quad e_2^\ast=0.
\end{equation}
Further, we can plug the fixed point solution into the RG equation for $g_1$ and $g_2$ to obtain 
\bea
	\frac{d g_1}{dl} &=& \left(\frac{3 (e_1^\ast - g_1) - 5 g_2}{24\pi^2 c^3} \right) g_1, \\
	\frac{d g_2}{dl} &=& \left(-1 + \frac{11e_1-13g_1-20g_2}{96\pi^2 c^3} \right) g_2.
\eea
The $g_2$ is already irrelevant at the tree level, it goes to zero $g_2^\ast =0$ at low energies. 
On the other hand, $g_1$ satisfies the RG equation $ \frac{d g_1}{dl} = \frac{3 (e_1^\ast - g_1)}{24\pi^2 c^3} g_1 $, so it goes to $g_1^\ast = e_1^\ast$. 
Within the fixed point obtained at the long-time limit, one can check the renormalization of the light velocity $c$ and find that $\frac{d c}{dl} =0$, consistent with our previous assumption.

\subsection{d=2+1} 

We present our preliminary investigation for the 2+1 dimension case. 
QED in 2+1 dimension ($\text{QED}_3$) is particularly interesting for its unique features compared to $\text{QED}_4$, such as confinement, chiral symmetry breaking, and dynamical mass generation~\cite{PhysRevLett.60.2575, pennington1991masses,PhysRevLett.62.3024,PhysRevD.52.6087,PhysRevD.54.4049}. 
It has been interpreted as an effective theory to describe various exotic physical phenomena like high-temperature superconductors~\cite{PhysRevB.66.094504,PhysRevB.66.054535}. 
To get an insight into the quenched QED in 2+1 dimension, we first solve the fixed point. Again, we assume a constant light velocity $c$, and one can solve the RG equations for $d=3$ and obtain two nonthermal fixed points solutions: $e_1=e_2=g_1=g_2=0$ and $e_2=6c^3\pi^2, e_1=-g_1, g_2=\frac{6}{5}(e_1+9c^3\pi^2)$. 
Only the trivial fixed point is stable for the initial physical isotropic interaction value $e_1=e_2=g_1=g_2$. 
The numerical results show that in the prethermal region the velocity $c$ flows to zero, while the Dirac fermion decouples from the gauge field with $e_{1;2}=g_{1;2}=0$. 
However, before the decoupling, the interaction terms flow to large finite values which is beyond the perturbation condition. 
On a long timescale $\Lambda t>\frac{1}{\Omega_0}$, since $c$ remains zero, the non-thermal fixed point is still trivial, while the equilibrium fixed point in $d=2+1$ should be finite. 
Thus, one can conclude that the quench will trivialize the quantum electrodynamics of the Dirac fermion in 2+1 dimension. 
Though all the interaction terms and the light velocity $c$ at the fixed point are zero, the fermion anomalous dimension at the nonthermal fixed point is zero. 
Note that, in contrast to the equilibrium $\text{QED}_3$, the fermion mass here is generated by the initial quench rather than the interactions, which distinguishes the postthermal behaviors from the regular $\text{QED}_3$. Considering higher-order corrections and taking the large-N limit may provide more insight into the 2+1d quenched $\text{QED}$.
\begin{figure}
	\includegraphics[width=8cm]{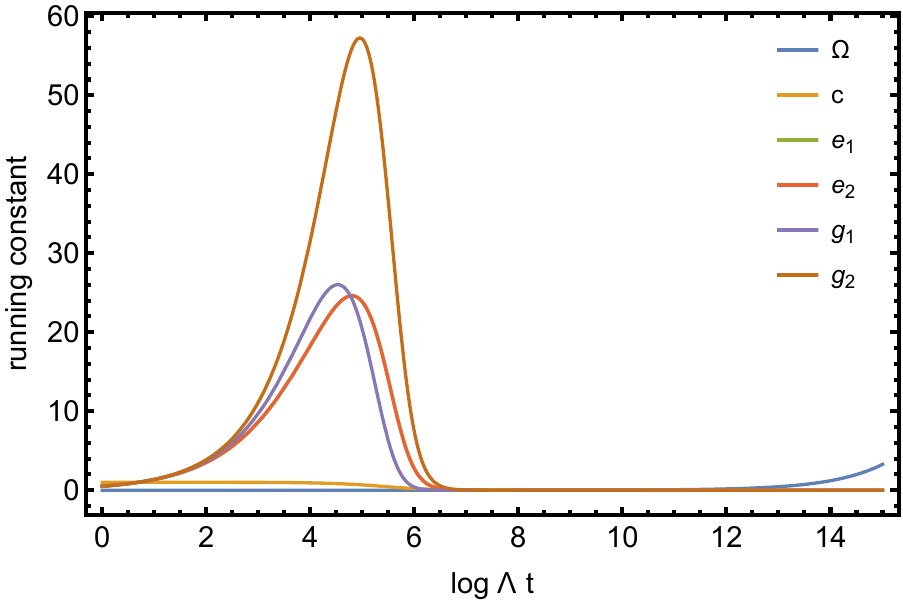}
\caption{Running coupling constants in three dimensions. 
There exists a prethermalization regime where interactions increase and seem to be relevant as the regular QED behavior in 2+1 dimension. 
On the long timescale, the light velocity flows to zero and there is a new nonthermal fixed point that is trivial $g_1=g_2=e_1=e_2=0$. 	\label{fig:3d}} 
\end{figure}

\end{document}